\documentclass[twocolumn,showpacs,preprintnumbers,amsmath,amssymb]{revtex4}

\usepackage{graphicx}
\usepackage{dcolumn}
\usepackage{bm}

\begin{document}


\title{Effect of the Zero-Mode Landau Level on Interlayer Magnetoresistance in Multilayer Massless Dirac Fermion Systems}

\author{Naoya Tajima$^1$}
\author{Shigeharu Sugawara$^2$}%
\author{Reizo Kato$^1$}%
\author{Yutaka Nishio$^2$}%
\author{Koji Kajita$^2$}%

\affiliation{%
$^1$RIKEN, Hirosawa 2-1, Wako-shi, Saitama 351-0198, Japan \\
$^2$Department of Physics, Toho University - Miyama 2-2-1, Funabashi-shi, Chiba 274-8510, Japan
}%

\date{\today}

\begin{abstract}
We report on the experimental results of interlayer magnetoresistance in multilayer massless Dirac fermion system $\alpha$-(BEDT-TTF)$_2$I$_3$ under hydrostatic pressure and its interpretation. We succeeded in detecting the zero-mode Landau level (n=0 Landau level) that is epected to appear at the contact points of Dirac cones in the magnetic field normal to the two-dimensional plane. The characteristic feature of zero-mode Landau carriers including the Zeeman effect is clearly seen in the interlayer magnetoresistance. 
 \end{abstract}

\pacs{71.10.Pm, 72.15.Gd}

\maketitle

Novoselov {\it et al.} and Zhang {\it et al.} have experimentally demonstrated that graphene is a zero-gap system with massless Dirac particles \cite{rf:1, rf:2}. They discovered new steps of the quantum Hall effect attributable to a zero-gap energy structure with linear dispersion. However, graphene is not a bulk crystal but a single layer of graphite. The bulk crystal zero-gap system was realized in the organic conductor $\alpha$-(BEDT-TTF)$_2$I$_3$ under high pressure \cite{rf:3, rf:4, rf:5, rf:6, rf:7, rf:8}. The existence of the massless Dirac fermions state in bulk crystal has already been discovered in graphite \cite{rf:9, rf:10} and Cd$_{\rm 1-x}$Hg$_{\rm x}$Te \cite{rf:11}. But graphite is not a zero-gap system but a semimetal. On the other hand, Cd$_{\rm 1-x}$Hg$_{\rm x}$Te is a three-dimensional (3D) semiconductor with narrow energy gap of more than 6 meV. In this sense, $\alpha$-(BEDT-TTF)$_2$I$_3$ is the first bulk 2D material with zero-gap energy bands. Thus, $\alpha$-(BEDT-TTF)$_2$I$_3$ provides us with a testing ground for the transport of the multilayer massless Dirac fermions system. In this Letter, we demonstrate that the characteristic feature of transport in multilayer Dirac fermions systems is clearly seen in interlayer magnetoresistance. 

A crystal of $\alpha$-(BEDT-TTF)$_2$I$_3$ consists of conductive layers of BEDT-TTF molecules and insulating layers of I$_3^-$ anions \cite{rf:12}. Since the conductive layers are separated by the insulating layers, carriers in this system have a strong 2D nature. The ratio of the in-plane conductivity to the interlayer conductivity is about $10^3$. Therefore, each conductive layer is expected to be almost independent. Under ambient pressure, this material behaves as a metal down to 135 K where it undergoes phase transition to a charge-ordered insulator. In the low temperature phase, a horizontal charge stripe pattern for +1 $e$ and 0 is formed \cite{rf:13,rf:14,rf:15,rf:16}. When the crystal is subjected to high hydrostatic pressure above 1.5 GPa, the metal-insulator transition is suppressed and the metallic region is preserved down to low temperatures \cite{rf:17, rf:18, rf:19}. This change in electronic state is accompanied by the disappearance of the charge ordering state, as shown by Raman experiments \cite{rf:16}. Note that in the high pressure phase, the resistance is almost constant over the temperature range of 300 K to 2 K. The conductivity for each conductive layer is estimated to be nearly equal to $e^2/h$. In contrast with the resistivity that is almost constant, carrier density and mobility exhibit strong temperature dependence. Carrier density ($n$) estimated by the Hall effect is proportional to the square of temperature ($T$) as $n\propto {T^{2}}$ over a wide temperature range. Such a peculiar character of this material remains an issue to be resolved.

Recent band calculations by Kobayashi {\it et al.} suggested that $\alpha$-(BEDT-TTF)$_2$I$_3$ under high pressure is in a zero-gap state \cite{rf:7,rf:8}. The bottom of the conduction band and the top of the valence band contact each other at two points (We call them \lq\lq contact points\rq\rq) in the first Brillouin zone \cite{rf:7, rf:8}. The Fermi energy of the electron system locates just at the contact point. These results were also supported by first-principle band calculations \cite{rf:20}. 

In the picture of a zero-gap system, transport phenomena of this material are understood as follows \cite{rf:6}. Carrier density written as $n \propto T^2$ is a characteristic feature of 2D zero-gap conductors with the Fermi energy located at the contact point \cite{rf:4, rf:6}. Carrier mobility, on the other hand, is determined as follows. According to Mott's argument, the mean free path ($l$) of a carrier subjected to elastic scattering can never be shorter than the wavelength ($\lambda$) of the carrier, so that $l \geq \lambda $\cite{rf:21}. For the case that there scattering centers exist with high density, $l \sim \lambda$. As the temperature is decreased, the mean free path becomes long because the wavelength ($\lambda$) becomes long with decreasing energy of carriers. As a result, the mobility ($\mu$) of carriers increases as $\mu \propto T^{-2}$ in the 2D zero-gap system \cite{rf:4, rf:6}. Consequently, the resistivity per layer (sheet resistance $R_{\rm S}$) is given as $R_{\rm S} =  h/e^2$, which is independent of temperature \cite{rf:6}.

One problem that remains to be explained is the transport phenomena under a magnetic field \cite{rf:4, rf:17, rf:18}. As transport phenomena of this system are extremely sensitive to the magnetic field, the samples are expected to be clean. Both quantum oscillations and the quantum Hall effect, however, were not observed in the magnetic field of up to 15 T. This is explained by the fact that the electron system is not degenerated. The Fermi energy of this material locates at the contact point both in the absence and in the presence of a magnetic field. Therefore, electrons do not have the Fermi surface. This is one of the characteristic features of intrinsic zero-gap conductors. 

Another important difference of zero-gap conductors from conventional conductors is the appearance of a Landau level at the contact points when magnetic fields are applied normal to the conductive layer. This special Landau level is called the zero-mode level. In this work, we succeeded in obtaining evidence of the zero-mode Landau level in our material. The idea of this work is as follows: Since the energy of this level is $E_{\rm F}$ irrespective of field strength, the Fermi distribution function is always 1/2. This means that half of Landau states in the zero mode are occupied. Note that in each Landau level, there are states with density proportional to $B$. Thus, the magnetic field creates mobile carriers. 

This effect can be detected in the interlayer resistance under a transverse magnetic field, as pointed out by Osada \cite{rf:22}. In this field configuration, the interaction between the electrical current and the magnetic field is weak, because they are parallel. On the other hand, the magnetic field in this direction (interlayer) yields Landau carriers. Hence, the conductivity is primarily proportional to the carrier density. The effect of the magnetic field appears only through the change in the carrier density. 

To demonstrate the appearance of zero-mode Landau carriers, we examined the inter-layer resistivity of this zero-gap conductor in the magnetic field. The resistance was measured at temperatures down to 0.06 K and at magnetic fields up to 7 T. 

A sample to which four electrical leads were attached was encased in a Teflon capsule filled with pressure medium (Idemitsu DN-oil 7373). The capsule was set in a clamp-type pressure cell made of MP35N hard alloy, and hydrostatic pressure of up to 1.7 GPa was applied. Resistance measurements were carried out using a conventional dc method with the electrical current along the $c$-crystal axis, which is normal to the 2D plane.


\begin{figure}
\includegraphics[viewport = 0 260 700 650, scale=.38, clip]{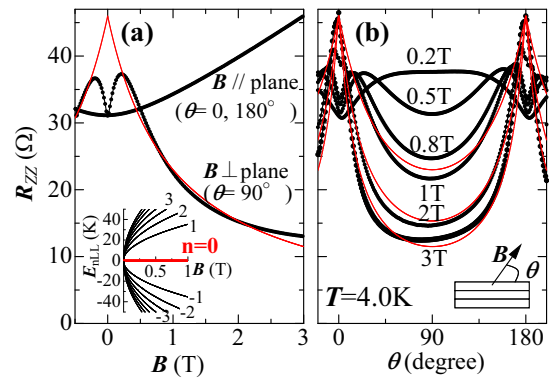}
\caption{\label{fig:1} (color online). (a) Magnetic field dependence of interlayer resistance under the pressure about 1.7 GPa at 4 K. When $B>0.2$ T, remarkable negative magnetoresistance is observed. As for negative magnetoresistance, fitting is done with Eq. (1), $R_{zz}\propto 1/(|B|+B_0)$ (red line). Inset shows the Landau levels of this system. (b) Angle dependence of magnetoresistance. When $\theta = 0$ and $\theta = 180^{\circ}$, the direction of the magnetic field is parallel to the 2D plane. $\theta = 90^{\circ}$ is the direction normal to the 2D plane. Fitting of the data measured at 1, 2,and 3 T was done using Eq. (1). }
\end{figure}

Figure 1 shows the magnetic field and the angle dependences of the inter-layer magnetoresistance. Resistivity behavior is dependent on the direction of the magnetic field. In the magnetic field parallel to the 2D plane ($\theta=0^{\circ}$), a gradual increase in resistivity was observed. In the field normal to the 2D plane ($\theta=90^{\circ}$), resistivity also increases at low fields below 0.2 T. Then, it decreases. Note that the resistivity drop is large: At 3 T, the resistivity is about 1/3 of that at 0.2 T. 

To understand the data for $B$ $\bot$ 2D plane in Fig. 1, we should examine zero-mode carriers quantitatively. The energy of Landau levels in zero-gap systems is expressed as $E_{\rm nLL}=\pm \sqrt{2e\hbar v_{\rm F}^2|{\rm n}|B}$, where $v_{\rm F}$ is the Fermi velocity and n is the Landau index \cite{rf:23}. In drawing the inset in Fig. 1 (a), we used $v_{\rm F}=10^7$ cm/s estimated from the data of the temperature dependence of carrier density. The carrier density depends on temperature as $n = \alpha T^2$, where $\alpha=(\pi^2k_{\rm B}^2/6c\hbar^2 v_{\rm F}^2)$ and $c$ is the lattice constant along the direction normal to the 2D plane \cite{rf:6}. According to band calculations, however, the Dirac cone in this system has highly anisotropic Fermi velocity $v_{\rm F}(\phi) $\cite{rf:7, rf:8}. Therefore, $v_{\rm F}=10^{7}$cm/s is an average of $v_{\rm F}(\phi)$. The inset in Fig. 1 (a) agrees with the realistic theory by Goerbig {\it et al.} \cite{rf:24}.

Since the experiments were performed at temperatures below 10 K, the energy gap between the zero-mode level and the n=1 level could exceed this temperature at rather low magnetic fields. At 1 T, for example, the energy gap is about 20 K. Therefore, if the experiment were performed at temperatures below 20 K and at magnetic fields above 1 T, most of the mobile carriers would be in the zero-mode Landau level. Such a situation is called the \lq\lq quantum limit \rq\rq. The quantum limit region is expressed as $k_{\rm B}T < E_{\rm 1LL}$. 

Carrier density in the quantum limit is expressed as $D(B)=B/2\phi_{0}$, where $\phi_{0}=h/e$ is the quantum flux. Factor 1/2 is the Fermi distribution function at $E_{\rm F}$. In moderately strong magnetic fields, the density of carriers induced by the magnetic field can be very high. At 3 T, for example, the density of zero-mode carriers will be 10$^{15}$ m$^{-2}$. This value is by about 2 orders of magnitude higher than the density of thermally excited carriers at 4 K, and, in the absence of the magnetic field, it is 10$^{13}$ m$^{-2}$. Therefore, carrier density in the magnetic field is expressed as $B/2\phi_{0}$ except for very low fields. This large change in the density of zero-mode carriers is the origin of the negative magnetoresistance, as shown in Fig. 1. 

Recently, Osada gave an analytical formula for interlayer magnetoresistance in a multilayer Dirac fermion system as follows:
\begin{equation}
\rho_{zz} = \displaystyle \frac{A}{\displaystyle |B_z|\exp\left[-\frac{1}{2}\frac{ec^2(B_x^2+B_y^2)}{\hbar|B_z|}\right]+B_0}, \label{eqn:1}
\end{equation}
where $A=\pi\hbar^3/2Ct_c^2ce^3$ is a parameter that is considered to be independent of the magnetic field if the system is clean. $B_0$ is a fitting parameter, $C$ is defined by $C=\int \rho_0(E)(-{\rm d}f/{\rm d}E){\rm d}E$ using the spectral density of the zero-mode Landau level and $\rho_0(E)$ satisfies $\int \rho_0 (E) {\rm d}E=1$ \cite{rf:22}. Note that only lattice constant $c$ is a material parameter. 

Except for narrow regions around $\theta=0^{\circ}$ and 180$^{\circ}$, this formula can be simplified to $\rho_{zz}=A/(|B_{z}|+B_{0})$. Using this formula and assuming $B_{0}$=0.7 T, we tried to fit the curves in Fig. 1. This simple formula reproduces well both the magnetic field dependence and the angle dependence of the magnetoresistance at magnetic fields above 0.5 T as shown by solid lines in Figs. 1 (a) and (b), which evidences the existence of zero-mode Landau carriers in $\alpha$-(BEDT-TTF)$_2$I$_3$ at high pressures.

Here, we briefly mention the origin of positive magnetoresistance around $\theta=0^{\circ}$ or 180$^{\circ}$. In the magnetic field in these directions, the Lorentz force works to bend the carrier trajectory to the direction parallel to the 2D plane. It reduces the tunneling of carriers between neighboring layers so that the positive magnetoresistance is observed. Note that the formula (1) for $\theta=0^{\circ}$ or 180$^{\circ}$ dose not correctly evaluate the effect of the Lorentz force and, thus, loses its validity. The value of the resistance peak depends weakly on the azimuthal angle. At 3 T, for example, the ratio of maximum value to the minimum value is less than 1.3. According to the calculation of interlayer magnetoresistance by Morinari, Mimura and Tohyama, the effect of Dirac cones with highly anisotropic Fermi velocity is averaged and gives rise to this small difference \cite{rf:25}.

An apparent discrepancy of the data from the formula (1) is also seen at both low and high magnetic fields normal to the 2D plane, because the model is oversimplified. Equation (1) was derived based on the quantum limit picture in which only the zero-mode Landau level is considered. In fact, each Landau level has a finite width due to scattering. At a sufficiently low magnetic field, the zero-mode Landau level overlaps with other Landau levels. In such a region, the formula (1) loses its validity. We can recognize this region in Fig. 1 below 0.2 T, where positive magnetoresistance is observed. This critical magnetic field shifts to a lower field with decreasing temperature, as shown in Fig. 2 (a).
 
\begin{figure}
\includegraphics[viewport = 0 300 700 650, scale=.38, clip]{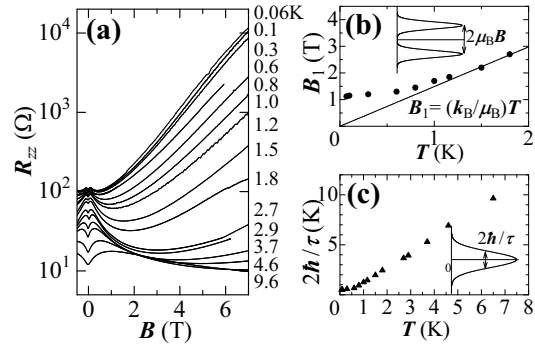}
\caption{\label{fig:2} (a) Magnetic field dependence of interlayer resistance for the magnetic field normal to the 2D plane at several temperatures from 0.06 K to 9.6 K. (b) Temperature dependence of $B_1$ when resistance is assumed to be $R_{zz} \propto \exp (B/B_1)$. (c) Temperature dependence of the width of Landau levels ($2\hbar / \tau$).}
\end{figure}

The deviation of data in the high field region is much more serious. In this region, the resistance increases exponentially with increasing field. This phenomenon is understood as follows.

In the above discussion, we did not consider the Zeeman effect. The Zeeman effect, however, should be taken into consideration because it has a significant influence on the transport phenomena at low temperatures. In the presence of a magnetic field, each Landau level is split into two levels with energies $E_{nLL}\pm \Delta E$, where $\Delta E$=$\mu_{\rm B}B$ is the Zeeman energy. This change in the energy structure gives rise to a change in the carrier density in Landau levels. In particular, the influence on the zero-mode carrier density is the strongest, because the energy level is shifted from the position of the Fermi energy. The value of the Fermi distribution function varies from $f(E_{\rm F})=1/2$ to $f(\Delta E)=1/(\exp(\Delta E /k_{\rm B}T)+1)$. At low temperatures where $k_{\rm B}T < \Delta E$, this effect becomes important. It works to reduce the density of zero-mode carriers and, thus, increases the resistance. 

The Zeeman energy when $B$ = 1 T is about 1 K. Therefore, in the experiment performed at 1 K, the deviation of experimental results from Eq. (1) is expected to start around 1 T. This is confirmed in Fig. 2 (a). At 1.8 K, for example, the deviation is prominent in fields above 2 T. This critical field shifts to about 0.3 T at 0.06 K. 

A definite evidence showing that the anomalous increase in the resistance at high field in Fig. 2 (a) is due to the Zeeman effect is given by examining the slopes of the curves at high fields. In this region, the magnetic field dependence of resistance is expressed as $R_{zz}\propto \exp (B/B_1)$, where $B_1$ is a parameter that depends on temperature. $B_1$ for $T$ =1.8 K is estimated to be about 2.7 T. At 0.06 K, it decreases down to about 1.1 T. In Fig. 2 (b), we plot the temperature dependence of $B_1$. Above 1 K, the $T$ vs $B_1$ curve is close to a line $B_1 =k_{\rm B}T/\mu_{\rm B}$. This is strong evidence that the Zeeman splitting of zero-mode Landau levels is the origin of the resistance that obeys the exponential law as $R_{zz} \propto \exp(2\mu_{\rm B}B/2k_{\rm B}T)$. In contrast, the behavior of $B_1$ below 1 K is understood in terms of the width of Landau levels. According to the theory of Osada, $\mu_{\rm B}B/(\hbar/\tau)\sim 1$ at the magnetoresistance minimum \cite{rf:22}. By using this relationship, the energy width of zero-mode Landau levels ($2\hbar / \tau$) is estimated as shown in the inset in Fig. 2 (c). With decreasing temperature, the energy width decreases linearly down to about 0.7 K, where it almost saturated. At 0.06 K, it is estimated to be about 0.5 K. This energy is much higher than the thermal energy. This is the Dingle temperature in the present system.

In conclusion, we provided ample evidence that $\alpha$-(BEDT-TTF)$_2$I$_3$ under high hydrostatic pressure is an intrinsic zero-gap conductor with the Dirac-type energy band. This is the first work that investigated the effect of the zero-mode Landau level on the transport phenomena of bulk zero-gap conductors. We succeeded in detecting the zero-mode Landau level that appeared at the contact points of Dirac cones under the magnetic field normal to the 2D plane. The large negative magnetoresistance observed in the interlayer resistance in longitudinal magnetic fields is due to the increase of the zero-mode carriers. Good agreement between experiment and theory was obtained. At high magnetic field, spin splitting of the zero-mode level becomes important, and the resistance obeys the exponential law. This is the characteristic feature of transport in multilayer zero-gap conductors. In graphite systems, a monolayer sample (graphene) is inevitable to realize a zero-gap state. On the other hand, in the present work, we demonstrate that bulk crystals can be 2D zero-gap systems. $\alpha$-(BEDT-TTF)$_2$I$_3$ provides us a suitable testing ground for the transport of the multilayer massless Dirac fermions system.

We thank Professor T. Osada, Dr. A. Kobayashi, Dr. S. Katayama, Professor Y. Suzumura, Dr. R. Kondo, Dr. T. Morinari, Professor T. Tohyama, and Professor H. Fukuyama for valuable discussions. This work was supported by Grants-in-Aid for Scientific Research (Nos. 19540387 and 16GS0219) from the Ministry of Education, Culture, Sports, Science and Technology, Japan.


\end{document}